\documentclass[showpacs,aps]{revtex4}
\usepackage{graphicx}
\begin{document}

\title{Relativistic center-vortex dynamics of a confining area law  }
\author{John M. Cornwall\footnote{Email cornwall@physics.ucla.edu}}
\affiliation{Department  of Physics and Astronomy, University of California, Los Angeles CA 90095}

\begin{abstract}
\pacs{11.15.Tk, 11.15.Kc \hfill UCLA/03/TEP/11}
 We  offer a physicists' proof that center-vortex theory requires the area in the Wilson-loop area law to involve an extremal area. This means that unlike in string-theory-inspired models of confining flux tubes,   area-law dynamics is determined  by integrating over Wilson loops only, not over surface fluctuations for a fixed loop. Fluctuations leading to to perimeter-law corrections  come from loop fluctuations as well as integration over finite-thickness center-vortex collective coordinates. In  $d=3$ (or $d=2+1$) we exploit a contour form of the extremal area in isothermal which is  similar to $d=2$ (or $d=1+1$) QCD in many respects, except that there are both quartic and quadratic terms in the action.    One major result is that at large angular momentum $\ell$ in $d=3+1$ the center-vortex extremal-area picture yields a linear Regge trajectory with Regge slope--string tension product $\alpha'(0)K_F$ of  $1/(2\pi )$, which is the canonical Veneziano/string value.  In a curious effect traceable to retardation, the quark kinetic terms in the action vanish relative to area-law terms in the large-$\ell$ limit, in which light-quark masses $\sim K_F^{1/2}$ are negligible. This corresponds to string-theoretic expectations, even though we emphasize that the extremal-area law is not a string theory quantum-mechanically.  We show how some quantum trajectory fluctuations as well as non-leading classical terms for finite mass yield corrections scaling with $\ell^{-1/2}$.   We compare to  old semiclassical calculations of relativistic (massless) $q\bar{q}$ bound states at large $\ell$, which also yield asymptotically-linear Regge trajectories, finding agreement with a naive string picture (classically, not quantum-mechanically) and disagreement with an effective-propagator model.   We show that contour forms of the area law can be expressed in terms of Abelian gauge potentials, and relate this to old work of Comtet.  
\end{abstract}

\maketitle

\section{Introduction}
\label{intro} 

The most familiar and elementary concept of confinement in QCD is the area law for Wilson loops.  For non-relativistic problems, such as the low-lying bound states of heavy quarks, the simple implementation of this idea is through a linearly-rising potential.  But after thirty years of QCD it is still not at all clear how to address, in the continuum and for $d\neq 2$, the dynamics of light-mass $q\bar{q}$ bound states and similar relativistic problems where the notion of potential breaks down.  In the early days of QCD, many authors (for example, \cite{co77}) suggested that a good approximation in $d=4$ would be to use an effective gluon propagator behaving like
\begin{equation}
\label{gluprop}
\Delta_{\mu\nu}(k^2)=(-g_{\mu\nu}+\cdots)\frac{ 8\pi^2K_F}{k^4}
\end{equation} 
for small momentum $k$, where $K_F$ is the string tension (the omitted terms in Eq. (\ref{gluprop}) are longitudinal).  The non-relativistic potential derived from this effective propagator is indeed linearly-rising, and in the relativistic regime yields a linearly-rising Regge trajectory at large angular momentum $\ell$.  (As discussed in \cite{co77}, the $q\bar{q}$ problem is relativistic for quark mass $M$ if $M^2\ll K_F\ell$.)  But this effective low-momentum propagator cannot be entirely correct; for example, it cannot be used as an effective propagator between gluons themselves, which are screened and not confined.

When phenomenological approaches such as \cite{co77}, using Eq. (\ref{gluprop}),  were first taken up, there was little understanding of how confinement actually worked in QCD.  Since then, many people have come to believe in the center-vortex picture of confinement.  (For early work in the center-vortex picture, see \cite{th78,c79,th79,mp79,no79}; an up-to-date review is in \cite{green}.) In this picture, center vortices (of co-dimension 2) can link topologically with a Wilson loop; with a spaghetti-like condensate of center vortices, the vacuum expectation value (VEV) over gauge fields of a Wilson loop results in an area law (for group representations where the center is non-trivially represented).  The vortices have finite thickness, but as far as the area law is concerned, thickness effects represent perimeter corrections ignored in this paper. 

The purpose of the present paper is twofold:  First, to argue, in physics language, that center-vortex dynamics leads to the prescription that, for a given Wilson loop, the area in the area law is extremal for that loop.  
Our second goal is to formulate the relativistic large-$\ell$ dynamics of quarks in this extremal-area center-vortex picture, and to compare this dynamics to earlier work, such as the effective-propagator model and various string models, in particular that of  \cite{cktwz}).    We will find that the extremal-area picture agrees (to the extent that they can be compared) with \cite{cktwz}, but gives a different value for the Regge slope than does \cite{co77}.
 We will study mostly semi-classical dynamics, which should be appropriate for large angular momentum $\ell$, and indicate how integration over fluctuations of Wilson loops gives rise to corrections which vanish at large $\ell$.  We will not address the full complications of relativistic $q\bar{q}$ dynamics, including quark spin, perturbative gluon exchange, and the like.  Our focus will be on finding the asymptotic slope of the linearly-rising Regge trajectory which (we will show) arises in the extremal-area picture.  

Even before what we will call QCD-inspired models, such as that of \cite{co77}, many authors had studied linearly-rising Regge trajectories in the context of the Veneziano model and various simple models of particles joined by strings; for example, see \cite{cktwz}.  The canonical relation between the string tension $K_F$ and the Regge slope $\alpha'(0)$, based on making classical string theory agree with the Veneziano model, is:
\begin{equation}
\label{veneziano}
\alpha'(0)K_F=\frac{1}{2\pi}.
\end{equation}
Other models give somewhat different values for the numerical constant on the right-hand side.  More recently, string theorists \cite{polstrom} have constructed models of $d=4$ strings which could represent hypothetical strings joining quarks.  There is no reason that  the general QCD-inspired model should give the Veneziano/string theory Regge slope, since the Veneziano model,  which pre-dates QCD, does not necessarily capture the true dynamical features of QCD, and there is a big difference between the properties of literal, physical strings which join quarks and the area-law dynamics proposed in the present paper, based on center vortices.  Moreover, to the extent that string theory is conformally-invariant there is no room in the string-theory action for separate kinetic terms for quarks.  (In an effective string theory, massive quarks can be represented as the ending of open strings on finitely-separated branes, yielding appropriate kinetic terms.)  The difference between literal strings and center-vortex dynamics is that the area in the area law for literal strings can fluctuate even though the contour (Wilson loop) is fixed, and these fluctuations need to be taken into account, while for center-vortex dynamics the area in question is minimal, and there are no surface fluctuations for a fixed bounding contour.

In Sec. \ref{arearg}  we discuss at some length the question of how one argues that the Euclidean center-vortex area law involves extremal areas.  The standard naive argument leads to an area law, but it is far from clear what surface and area are to be used for a given Wilson loop.  We adopt a center-vortex model which retains only those features necessary to discuss this question of the area law.  The model consists of a condensate of infinitesimally-thin closed vortex lines ($d=3$) or sheets ($d=4$) a finite fraction of which have infinite length.  These vortices  are characterized by two lengths:  A persistence length and a density of vortex pierce points per unit area.  In this model we show that the area in the area law is indeed extremal.  As stated above, finite-thickness vortices lead only to perimeter-law corrections.

Once the extremal-area law is established, we face the problem (the Plateau problem) \cite{cour} of finding the extremal area associated with a given bounding Wilson loop.
In Sec. \ref{hope} we remind the reader of some well-known obstacles to solving the Plateau problem.   These include the behavior of extremal areas for poorly-behaved contours, such as space-filling contours, and the general existence of more than one extremal area for a given contour.  We speculate that some of these peculiarities might actually be observable in QCD dynamics.   For example, two parallel and equal circles separated axially have (at least) two extremal areas:  The areas of the circles, and a catenoid joining them.  One or the other is a global extremum, depending on the axial separation of the circles.  The transition from one global minimum to the other may be related to the behavior of the adjoint potential, which is a breakable string.

In Sec. \ref{contour} we consider how to express a Euclidean extremal area in contour form.  This is actually elementary but not commonly used (string theorists, for example, usually use standard integrals over the world sheet).   The easiest case is $d=2$, where any simple area is a global minimum, and can be expressed as a quadratic form in the contour.  In higher dimensions the area law is simplest in Dirichlet form, which requires imposition of gauge conditions in the form of choosing isothermal coordinates.  The Dirichlet form of the area is still quadratic in the variables describing the surface, but this is misleading, because this is so only in isothermal coordinates.  We show that imposing isothermality leads to quartic terms in the surface variables, in any dimension greater than two.  (In $d=2$ isothermality is automatic.)  After expressing the Dirichlet form of the extremal area as a contour functional, we take up the issue of how to enforce isothermality, and end up with a modified contour functional which is quartic in the contour variables.  In certain cases these quartic terms may be dealt with as perturbations around a flat surface.  This area contour functional, plus kinetic terms, describes a one-dimensional field theory of contours.  This is to be contrasted to string-theoretic forms of confinement, which lead to two-dimensional field theories. It is far from easy (but not impossible; see \cite{polstrom}) to find the right form of such a theory for confinement, since standard bosonic string theory has a critical dimension of 26.  

In $d=3$ isothermality can be identically satisfied by a choice of functions describing the surface, as Weierstrass and Enneper showed long ago.  However, the Enneper-Weierstrass formulation does not make it easy to express an extremal area as a functional of its bounding contour, in general.  As would be expected from the paragraph above, the Enneper-Weierstrass area form is quartic in its variables.  

We also show in Sec. \ref{contour} that the Dirichlet area law in contour form has an interpretation, in $d$ dimensions, as a contour integral over $d$ Abelian gauge potentials.  Alternatively, in $d=3$ or $d=2+1$ one may choose a single Abelian gauge potential and a complex scalar field as the degrees of freedom.  These, as Comtet \cite{comtet} as shown, can be identified with Witten's \cite{wit} degrees of freedom for spherically-symmetric $d=4$ $SU(2)$ gauge theory.  We will pursue this connection further elsewhere.

 In Sec. \ref{appl} we apply this formalism to some simple problems of (spinless) $q\bar{q}$ relativistic dynamics in $d=2$, $d=1+1$, and in Sec. \ref{3dsec} to the $d=3+1$ problem of \cite{co77}, \cite{cktwz} (which is, at the classical level, actually a problem in $d=2+1$, since one spatial coordinate is inert).   Given the action (including kinetic terms) as a functional of Wilson loops, one must integrate over all Wilson loops to construct Green's functions. An example of such an integral, for scalar ``quarks" in Euclidean space, is the free quark propagator:
\begin{equation}
\label{green}
\langle \phi (x)\bar{\phi}(y)\rangle = \int_0^{\infty}ds \int_x^y (dz) \exp [-\frac{M}{2}\int_0^{s}d\tau \dot{z}(\tau )^2+1)]
\end{equation}
where $M$ is the scalar-quark mass; $s,\tau$ are  proper-time variables; and the integral $(dz)$ is an integral over  contours, subject to the constraint that these contours pass through  $x$ at $\tau =0$ and $y$ at $\tau = s$.   For $q\bar{q}$ dynamics one introduces a quark and and anti-quark propagator such as in Eq. (\ref{green}), plus the appropriate area-law action, and integrates over both quark and anti-quark contours.  As one expects, the $d=2$ or $d=1+1$ dynamics is exactly equivalent to the well-studied and ancient problem of large-$N$ two-dimensional QCD, and our discussion of two dimensions is largely review.  

In this Section we  take up the problem of circular $q\bar{q}$ orbits at large angular momentum $\ell$, using the extremal-area formalism, and compare to the   results of \cite{co77}, which used the effective propagator of Eq. (\ref{gluprop}), and of \cite{cktwz}, which used a naive string model.      Very fortunately, it is quite elementary to find the extremal surface expressed in isothermal coordinates (a helicoid) and its area analytically,  corresponding to the semiclassical orbits (a double helix) arising from solving the classical equations of motion.  One surprising feature of the extremal-area model is that in the small-mass regime  (meaning $M^2\ll K_F\ell$) the quark kinetic energy $M\gamma$, where $\gamma$ is the usual relativistic factor, vanishes in comparison to the energy coming from the extremal-area action.  This is rather different from the results of \cite{co77}, where $M\gamma$ does not vanish relative to area-law energy.  The significance of this vanishing is that, in some sense, the extremal-area law is rather like string theory which has no separate kinetic terms for quarks.

Note that this small-mass regime does not require literal masslessness, which is in any case impossible in QCD, since even quarks of zero current mass undergo chiral symmetry breakdown driven by confinement itself, which
leads to mass generation with $M\sim K_F^{1/2}$.  We will not attempt to model such effects here, and will in fact ignore area-law terms responsible for constituent-mass generation.  But we do show that both non-leading classical terms and certain quantum corrections both yield additive corrections to the bound-state energy $E$ which are of the same form as mass terms, that is, $E=(2\pi K_F\ell )^{1/2}+O(K_F^{1/2})$.  

As a result of the vanishing of quark kinetic energy relative to area energy for the extremal-area model, the Regge slope--string tension product of this model is precisely that of the Veneziano/string model of Eq. (\ref{veneziano}).  This slope differs significantly from that found in \cite{co77}; we find that this is because for any area-like model where quark kinetic energy is significant in the large-$\ell$ limit, the Regge slope is necessarily smaller than the Veneziano/string slope.     We also show that identically the same classical area action governs the present work and that of \cite{cktwz}.  These authors quote, but do not derive, the Veneziano/string slope as applicable to their model.  The identity of classical actions does not mean that the string model of  \cite{cktwz} and the extremal-area model are {\em physically} identical, because presumably in the string model the world surface of the string is subject to the usual sort of quantum fluctuations that strings have, even  when the Wilson loop it joins is fixed. But in the center-vortex picture of the present paper the nature of the fluctuations is quite different:  For a given Wilson loop there is a unique surface, at least if the loop is sufficiently well-behaved, and perimeter-law fluctuations arise from the thickness of center vortices, not from surface fluctuations.

  In Section \ref{appl}  we also show how certain integrations over quantum fluctuations of the contours lead to corrections which vanish as $\ell\rightarrow \infty$.

\section{Argument for extremal area}
\label{arearg}

For simplicity, we restrict our explicit considerations  to $SU(2)$ gauge theory; this avoids complications with nexuses, baryonic Wilson loops, and similar topological elaborations occurring for $SU(N),\;N\geq 3$.  Our center-vortex model consists of a vacuum condensate of infinitesimally-thick oriented vortices, of co-dimension 2 (so points in $d=2$, closed lines in $d=3$, closed two-surfaces in $d=4$), characterized by two (presumably related) quantities:  An areal density $\rho$ of vortex pierce points per unit area and a persistence length $l$.   The density $\rho$ is the number of vortices penetrating an element of area, divided by that area, each pierce point being counted as +1.  Each vortex can be considered as a closed random walk which avoids itself and other vortices (subject to the constraint that the density is $\rho$), and the persistence length is, as usual, the length over which the direction of the random walk decorrelates.
  
As is well-known, the VEV of a Wilson loop is an average over vacuum gauge-field configurations of phase factors constructed from link numbers of the vortex configuration with the Wilson loop. For $SU(2)$ the phase factor is $\exp (i\pi L)$ for link number $L$, so that only the link number mod 2 counts.  The total link number of a given vortex with a Wilson loop is found by adding numbers $\pm 1$ (depending on the relative orientation of vortex and Wilson loop) for each oriented intersection of the vortex with a surface spanning the Wilson loop.   
Because the linkage of an idealized zero-thickness center vortex and a Wilson loop is topological, it is tempting to say that the area-law part of the Wilson-loop VEV is stationary under changes of the surface spanning the loop when the loop is held fixed, and so the surface is extremal.   This is, however, a seriously incomplete argument.  Let us review the standard but sloppy argument for the area law in center vortex theory.  We write the Wilson-loop VEV as:
\begin{equation}
\label{link}
\langle W\rangle =\langle \exp (i\oint dx_iA_i)\rangle =\langle e^{i\pi L}\rangle ;\;L=\sum^N L_i
\end{equation} 
where $L_i=\pm 1$ is the link number associated with the $i^{th}$ pierce point of the spanning surface, and $N$ is the total number of pierce points of vortices with the spanning surface.  Some vortices intersect the spanning surface only once (or an odd number of times) and contribute -1 to the sum of terms in the loop VEV; others intersect not at all  (or an even number of times) and contribute a phase factor +1, as if they were not present at all.

Suppose that the Wilson loop under consideration is large, in the sense that all its spatial scales are large compared to the persistence length $l$ and the density length $\rho^{-1/2}$, and mathematically benign, for example, $C^{\infty}$, not self-intersecting, and so forth.  Then the total number of pierce points $N=\rho A$ of the spanning surface $S$, with area $A$, is large.  Each pierce point is associated with a partial link number $L_i=\pm 1$.  A certain fraction $\xi$ of these $L_i$ are independent and random; the remainder are link numbers of vortices which are small compared to the Wilson loop scales, and are not linked. One might then argue that the Wilson-loop VEV is found by assuming Gaussian statistics for the independent link numbers.  (Actually, Poisson statistics, but this fine point is irrelevant to us).  Then this VEV would be given by: 
\begin{equation}
\label{gauss}
\langle e^{i\pi L}\rangle=\exp [\frac{-\pi^2}{2}\langle L_i^2\rangle N_{ind}]
\end{equation}
with $\langle L_i^2\rangle=1$.  Here $N_{ind}=\xi \rho A$ is the total number of independent vortices. 

At this point an area law is apparently established, but it is far from clear what area is to be used.
To say that the area $A(S)$ in the area law is the area of an arbitrarily-chosen surface $S$ makes no sense for the general spanning surface, for example, one whose scale lengths and area are astronomical in scale, although there is nothing wrong with using such a surface.  The VEV is quite independent of which surface is chosen, contradicting the naive interpretation of Eq. (\ref{gauss}). Yet there must be some surface $S_{WL}$, of area $A_{WL}$, such that the VEV is given by this equation:
\begin{equation}
\label{rightlaw}
\langle e^{i\pi L}\rangle=\exp [\frac{-\pi^2}{2}\langle L_i^2\rangle \xi \rho A_{WL}].
\end{equation}
  This must be the area of some surface $S$ spanning the Wilson loop; if $S$ did not span it, some vortex linkages would not be counted.  

What went wrong is the assumption that all the $L_i$ are actually independent, even for vortices very long compared to the Wilson loop size.  Consider the vortex associated with a particular pierce point $i$.  Let us consider a topological configuration consisting of two surfaces each spanning the Wilson loop, one of them being an arbitrary surface $S$ that is far from the correct surface and the other the correct (but unknown) surface $S_{WL}$.   The surface $S+S_{WL}$ is a closed surface with no boundaries; we choose it to be oriented.  Since all vortices are closed, if a vortex pierces the surface $S$ it must pierce the surface $S+S_{WL}$ an odd number of times more.  For simplicity, we state the argument explicitly for vortices which penetrate exactly once more.  Consider the fate of a given vortex which enters the volume enclosed by $S+S_{WL}$ by penetrating $S$. Upon exiting this volume it can penetrate either $S$ or $S_{WL}$ again.  If it penetrates $S$ but not $S_{WL}$ it is not linked to the Wilson loop; otherwise it is.  Evidently, the probability $P$ that the given vortex will in fact penetrate the Wilson loop is $P=A(S_{WL})/A(S)$.  In effect, the mean-square link number $\langle L_i^2\rangle $ is not 1, but $P$, and the area law does come out in the form of Eq. (\ref{rightlaw}), and is independent of the originally-chosen surface $S$.  Unfortunately, this argument does not tell us how to choose $S_{WL}$.

This argument holds unless the surface $S$ is very close to the correct surface $S_{WL}$, in which case the link numbers and pierce points of the two surfaces are strongly correlated.  
Consider a new choice of spanning surface $S_{WL}+\delta S$ which is very close to the correct surface, and calculate the Wilson-loop VEV with the surface $S_{WL}+\delta S$.  The area of this spanning surface is $A_{WL}+\delta A$. Either the link numbers $L_i$ of the correct surface $S_{WL}$  become those of the new surface, without change, or new pierce points  arise in pairs, with total link number 0.  These new pierce points arise if in going from $S_{WL}$ to $S_{WL}+\delta S$ an unlinked vortex which penetrated one of these surfaces twice does not penetrate the other one at all.  But these effects cancel when averaged over many gauge configurations,  and provided that the spatial scale of the deviation $\delta S$ is small compared to the persistence length, do not happen at all.  The result is that the Wilson-loop VEV is precisely the same for the two surfaces:
\begin{equation}
\label{extremal}
\langle e^{i\pi L}\rangle =\exp [-\frac{\pi^2}{2}\xi \rho A(S_{WL})]=\exp [-\frac{\pi^2}{2}\xi \rho (A(S_{WL})+\delta A)]
\end{equation}
and so $\delta A(S_{WL})=0$; the correct surface is extremal.

\section{Extremal surfaces and confinement:  A hopeless problem?}
\label{hope}

As mathematicians know, it is hopeless to give a completely general solution to the Plateau problem, because of the variety of contour and surface types that could be encountered. It is a theorem that any Jordan curve possesses at least one extremal area which spans the curve, but there can be infinitely many such areas, in pathological cases.  There is no particular reason, for example, to suppose that the contour spanned by a minimal surface is at all regular.  An example is a random walk, or something like Hilbert's space-filling curve.

It could happen that such a  curve  generates an area which grows faster than $R^2$, where $R$ is a maximal diameter of the contour.  From Eq. (\ref{diripart}) below, one proves by elementary arguments the inequality for the minimal area $A$:
\begin{equation}
\label{inequal}
A\leq \frac{1}{2}RL
\end{equation}
where $L$ is the length of the contour and $R$ a maximum diameter.  For a contour generated by a random walk, $L=R^2a^{-1}$ where $a$ characterizes the average step length for the walk.  It is then possible, from Eq. (\ref{inequal}), that the area grows as fast as $R^3$.  However, if this rate of growth were actually achieved, it could not be important for the infrared behavior of Wilson loops, since the contributions of a loop averaged over gauge fields decays exponentially as $\exp (-KA)$ where $K$ is the string tension.  The mathematical reality is that we can only make progress with well-behaved contours, and we assume good behavior at every step of our arguments.

\begin{figure}
\includegraphics[width=6in]{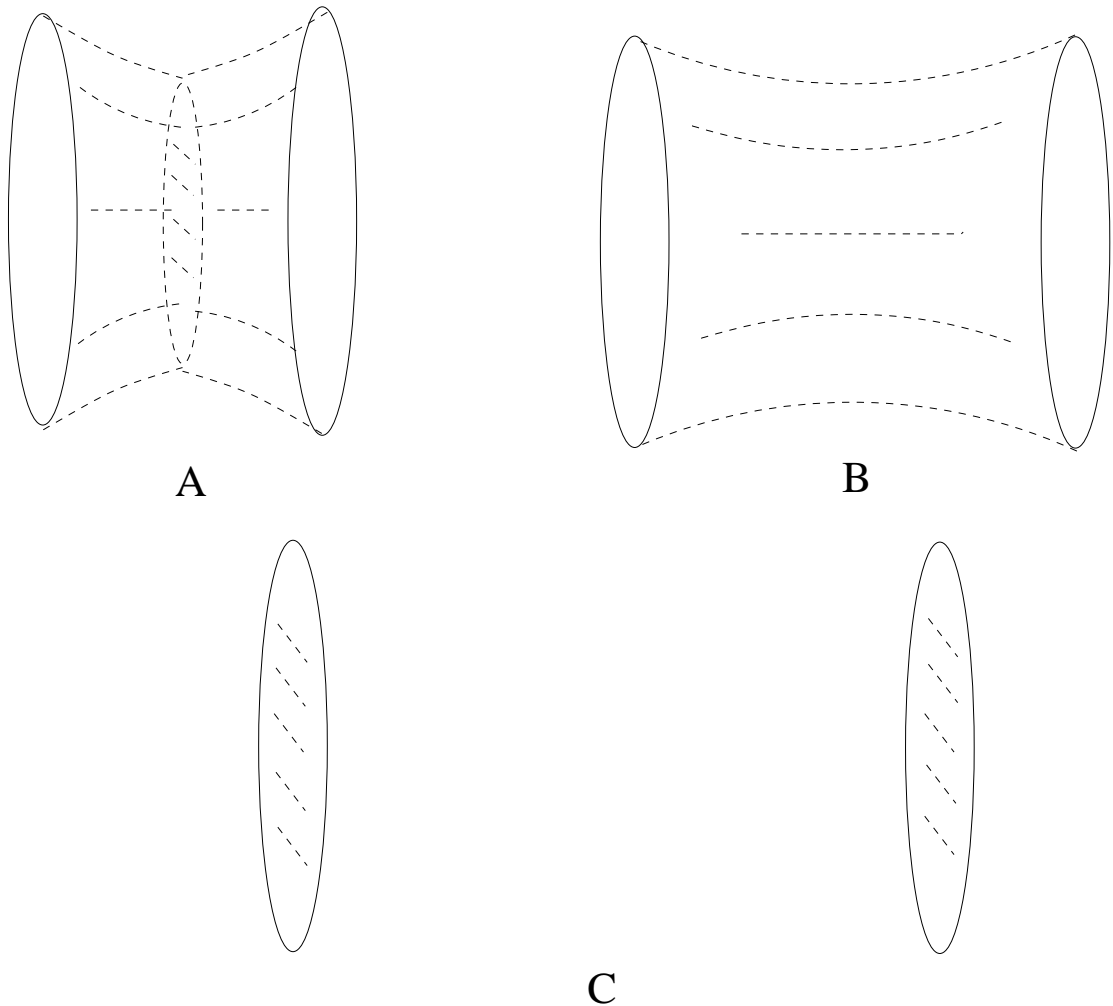}
\caption{\label{fig1} A.  Two parallel circles with extremal surfaces meeting at the Steiner angle.  B.  Two circles joined by a catenoid.  C.  If the circles are far enough apart, the global minimum surface is the two planar surfaces of the circles.}
\end{figure}

Even with this assumption of well-behaved contours and surfaces, some interesting questions arise.
There are many contours which have more than one extremal area.  Consider, for example, two equal-size circles whose planes are parallel; depending on the separation between the circles (in units of the circles' radius) several extremal areas are possible, as shown in Fig. \ref{fig1}.  We may think of the loops in Fig. \ref{fig1} as being idealized representations of nearby $q\bar{q}$ pairs as would occur in a hybrid meson.  The mere geometrical existence of an extremal area does not mean it makes sense in a confining gauge theory.  For example, the configuration of surfaces shown in Fig. \ref{fig1}A is not sensible in $SU(2)$ gauge theory, with the usual interpretation of the intersection of a center vortex (a closed line that, if linked $K$ times to a fundamental Wilson loop, gives rise to the factor $(-)^K$ for the Wilson loop).  This intersection number $K$ can be calculated by counting the number of signed intersections of the Wilson loop with an orientable surface spanning the loop.  A simple center vortex can be linked once positively to both Wilson loops of Fig. \ref{fig2}A, and then it contributes a factor of unity to the product of Wilson loops, just as if it were linked to neither.  But in Fig. \ref{fig2}A it is possible either for a vortex so linked to pass through the central plane surface and for it not to; in the former case, one would assign a factor of -1 to the product of Wilson loops, and in the latter, a factor of +1.  This figure can make sense for $SU(3)$ with the assignment of fluxes to surfaces consistent with the rules for associating a nexus world line with the line where three surfaces meet in the figure.  

Go back to $SU(2)$; both Figs. \ref{fig1}B,C make sense.  The same link numbers are found by counting penetrations of either spanning surface by a center vortex.  But the areas to be used in the area law for Wilson loop VEVs are clearly different, and so apparently is the interpretation.  For Fig. \ref{fig1}B the Wilson loops are correlated, in that the spanning surface depends on both loops, but in Fig. \ref{fig1}C they are uncorrelated.    Interestingly, as one would expect on physical grounds, when the separation of the loops is large enough (roughly equal to their common radius) the catenoid suggested by Fig. \ref{fig1}B is not a global minimum; only the independent planar surfaces of Fig. \ref{fig1}C are.  This suggests the possibility of {\em short-ranged} van der Waals-like forces between two hadrons, not necessarily associated with meson exchange, and also has implications for the behavior of the screening law for adjoint Wilson loops, and area laws for $N$-ality non-zero representations of $SU(N)$ for $N\geq 4$.  We will explore these issues elsewhere.

\section{Finding the contour form of extremal areas}
\label{contour}

Certain of our remarks apply in all dimensions,  while some are specific to $d=3$.  Also of interest are elementary area formulas in $d=2$, because their contour version resembles what we will find in higher dimensions to some extent, with the contours acting as currents sources for a $d=2$ Abelian gauge propagator.  Later we will show that such Abelian gauge propagators arise in all dimensions.  In another section we will review $d=2$ dynamics, which is governed by a quadratic and hence soluble action.  For $d>2$ quartic terms arise from the isothermality conditions.

The area of a surface is given by the usual expression:
\begin{equation}
\label{area1}
A=\int d^2\sigma \eta^{1/2}
\end{equation} 
where the $\sigma_i,\;i=1,2$ are coordinates on the surface and $\eta$ is the determinant of the induced metric:
\begin{equation}
\label{indmetric}
\eta_{ij}=g_{\alpha\beta}\frac{\partial x^{\alpha}}{\partial \sigma_i}\frac{\partial x^{\beta}}{\partial \sigma_j}.
\end{equation}
Greek indices run from 1 to $d$.
For now we will be concerned with a Euclidean embedding space, so $g_{\alpha\beta}=\delta_{\alpha\beta}$.  We do not need to distinguish covariant from contravariant indices, but for notational clarity we will always use contravariant notation for Greek indices and covariant notation for Latin indices.
If we impose the isothermal gauge conditions
\begin{equation}
\label{iso}
\eta_{ij}=\lambda(\sigma_a)\delta_{ij}:\;\;\frac{\partial x^{\alpha}}{\partial \sigma_1}\frac{\partial x^{\alpha}}{\partial \sigma_1}=\frac{\partial x^{\alpha}}{\partial \sigma_2}\frac{\partial x^{\alpha}}{\partial \sigma_2}=\lambda ;\;\;\frac{\partial x^{\alpha}}{\partial \sigma_1}\frac{\partial x^{\alpha}}{\partial \sigma_2}=0.
\end{equation}
 the equations for an extremal surface of area $A$ (defined as one with zero mean curvature everywhere) are derived from a simple Dirichlet action:   
\begin{equation}
\label{diri}
A_D=\frac{1}{2}\int d^2\sigma (\partial_ix^{\alpha})^2. 
\end{equation}
These equations  of an extremal surface are $\nabla^2x^{\alpha}$=0, which means  that the coordinates can be extended to holomorphic functions of the complex variable $\sigma\equiv \sigma_1+i\sigma_2$.
We write this extension of real coordinates $x^{\alpha}$ to complex coordinates $X^{\alpha}$:
\begin{equation}
\label{complex1}
X^{\alpha}=x^{\alpha}+iq^{\alpha}
\end{equation}
where $x^{\alpha},\;q^{\alpha}$ are real and the $X^{\alpha}$ are holomorphic functions of $\sigma$. The Cauchy-Riemann relations relate the real and imaginary parts of the derivatives: 
\begin{equation}
\label{cr}
\partial_ix^{\alpha}=\epsilon_{ij}\partial_jq^{\alpha},\;\partial_iq^{\alpha}=-\epsilon_{ij}\partial_jx^{\alpha}
\end{equation}
(with $\partial_i=\partial/\partial \sigma_i$).
By virtue of the Cauchy-Riemann relations the isothermality condition of Eq. (\ref{iso}) becomes:
\begin{equation}
\label{isothermal}  
(\partial_{\sigma}X^{\alpha})^2=0.
\end{equation}

In Eq. (\ref{diri}) we distinguish between the true area $A$ and the Dirichlet area $A_D$, because in general, if the isothermal conditions are not imposed, the most one can say is that $A\leq A_D$, as follows from:
\begin{equation}
\label{adiri}
A=\frac{1}{2}\int d^2\sigma \{[\sum_{\alpha}|\partial_{\sigma}X^{\alpha}|^2]^2-|\sum_{\alpha}(\partial_{\sigma}X^{\alpha})^2|^2\}^{1/2}.
\end{equation}
The Dirichlet area 
\begin{equation}
\label{area2}
A_D=\frac{1}{2}\int d^2\sigma |\partial_{\sigma}X^{\alpha}|^2.
\end{equation} 
agrees with the true extremal area when isothermality is imposed.  Note that one may write $A_D$ in a number of ways, for example,
\begin{equation}
\label{dirimulti}
A_D=\frac{1}{2}\int d^2\sigma (\partial_iq^{\alpha})^2
\end{equation}
as well as in the original form of Eq. (\ref{diri}). 

Two-dimensional area formulas illustrate these principles very easily.

\subsection{The area formula in two dimensions}

In $d=2$ Euclidean space every area associated with a piecewise once-differentiable non-self-intersecting closed contour $x_i(t)$ is minimal, so the required answer is just the usual area.

We label the two coordinates $x_i$ and choose $\sigma_i=x_i$ (thereby fulfilling the isothermal conditions).  In this case, Eq. (\ref{diripart}) becomes the usual formula:
\begin{equation}
\label{diri2}
A=\pm\frac{1}{2}\int dt \epsilon_{ij}x_i\dot{x}_j 
\end{equation}
where the contour is given as a function of a parameter $t$, and the overdot indicates the derivative with respect to $t$.  The area must be positive, of course, and the sign is chosen according to choices made about the direction of the contour.  

A second form, making clear the connection between $d=2$ QCD and an area law, is the $d=2$ version of Eq. (\ref{finalarea}) and it states the area law as an action involving a massless vector propagator.     This form is:
\begin{equation}
\label{2darea}
A=\oint dx_i \oint dx_j' \Delta_{ij}[x-x'];\;\;\Delta_{ij}=(\delta_{ij}+\cdots)(\frac{1}{2\pi})\ln |x-x'|.
\end{equation} 
(We allow for longitudinal terms, containing $\partial_i$ or $\partial_j$ and indicated by the dots; they contribute nothing to the area.) Of course, $\Delta_{ij}$ is the $d=2$ massless vector propagator, so that large-$N$ $d=2$ QCD is truly an area law, as is well-known.

One can show directly that  Eq. (\ref{2darea}) gives the usual area by using Stokes' theorem on each contour integral, plus the fact that the Laplacian on $\Delta_{ij}$ gives a delta function, to arrive at $A=\int d^2\sigma$.  

\subsection{Contour formulas in higher dimensions}

The contour form of the Dirichlet area $A_D$  is found by Stokes' theorem and the observation that
\begin{equation}
\label{stokes}
(\partial_ix^{\alpha})^2=\epsilon_{ij}\partial_j(q^{\alpha}\partial_ix^{\alpha}).
\end{equation}
One contour form of $A_D$ is then:
\begin{equation}
\label{diripart}
A_{Dc}=\frac{1}{2}\oint_{\Gamma} d\sigma^iq^{\alpha}\partial_ix^{\alpha} 
\end{equation}
where $\Gamma$ is the contour spanned by the minimal surface.  We give it a new name $A_{Dc}$ for reasons which will become clear later.

 The functions $q^{\alpha}$ are not immediately known in terms of the $x^{\alpha}$, but they can be eliminated by using the Cauchy-Riemann relations in the standard integral for an antiderivative $G$:
\begin{equation}
\label{antider}
G(\sigma )=\frac{-1}{2\pi i}\oint_{\Gamma} d\sigma'G'(\sigma')\ln (\sigma'-\sigma )+{\rm const.}
\end{equation}
where $G'=\partial G /\partial \sigma'$. The integral gives zero if $\sigma$ lies outside the contour $\Gamma$; the unspecified constant depends on how the contour is constructed.  

We write
\begin{equation}
\label{log}
\ln (\sigma' -\sigma )=\ln |\sigma'-\sigma |+i\gamma;\;\gamma =\arctan [\frac{\sigma'_2-\sigma_2}{\sigma'_1-\sigma_1}]
\end{equation}
and observe the Cauchy-Riemann relations: 
\begin{equation}
\label{equalder}
\epsilon_{ij}\partial_j\ln |\sigma'-\sigma |=\partial_i\gamma.
\end{equation}
It is then straightforward to integrate by parts twice in Eq. (\ref{antider}), keeping track of endpoint terms which can survive if the contour crosses the logarithmic branch line, and find:
\begin{equation}
\label{disprel}
G\equiv G_1+iG_2=\frac{1}{\pi}\oint d\sigma_i'\ln |\sigma'-\sigma |\epsilon_{ij}\partial_jG_1+\frac{i}{\pi}\oint d\sigma'_i\ln |\sigma'-\sigma |\partial_iG_1+{\rm const.}
\end{equation}
This equation allows us to express the $q^{\alpha}$, which are the unphysical imaginary parts of the surface coordinates and correspond to $G_2$ in Eq. (\ref{disprel}), as integrals over derivatives of the physical coordinates $x^{\alpha}$ (corresponding to $G_1$).  By expressing $q^{\alpha}$ in terms of $\partial_ix^{\alpha}$ through equations of the type of Eq. (\ref{disprel}) one finds that $A_{Dc}$ is given by:
\begin{equation}
\label{finalarea}
A_{Dc}=\frac{1}{2\pi}\oint d\sigma_i\partial_ix^{\alpha}(\sigma )\oint d\sigma'_j\partial_jx^{\alpha}(\sigma' )\ln |\sigma'-\sigma |.
\end{equation}

Eq. (\ref{finalarea}) is apparently a contour-integral form for the Dirichlet area.  However, more work is needed, because in Eq. (\ref{finalarea}) one can specify independently functions $x^{\alpha}(\sigma )$ and a contour on the world sheet $\sigma_i=\sigma_i(t)$, where the parameter $t$ labels points along the contour.  The contour in target space is $x^{\alpha}(t)\equiv x^{\alpha}[\sigma_i(t)]$. For a fixed target-space contour the area $A_{Dc}$ can take on a continuum of values.  Insuring that the contour area $A_{Dc}$ agrees with the Dirichlet area $A_D$ requires that isothermality be imposed.

\subsection{Dealing with isothermality}
\label{isotherm}

We now show that if the isothermal conditions are imposed that the contour form of the area, $A_{Dc}$, depends on the contour $\sigma_i(t)$ only to the extent that the original form of the Dirichlet area (that is, $A_D$ of Eq. (\ref{area2})) depends on the contour, when the functional dependence of $x^{\alpha}$ is held fixed. This means that for a variation $\delta\sigma_i$ of the $\sigma$ contour, the target-space contour changes by:
\begin{equation}
\label{constraint}
\delta x^{\alpha}(t)=\delta \sigma_i(t)\partial_ix^{\alpha}(\sigma_i).
\end{equation}
The contour variation of the Dirichlet area $A_D$ is:
\begin{equation}
\label{contourvar}
\delta A_D=\oint \epsilon_{ij}d\sigma_i\delta\sigma_j(\partial_kx^{\alpha})^2.
\end{equation}
As for $A_{Dc}$ one finds, by using the Cauchy-Riemann relations, its variation:
\begin{equation}
\label{varadc}
\delta A_{Dc}=2\oint d\sigma_i\delta \sigma_k\epsilon_{il}\partial_lx^{\alpha}\partial_kx^{\alpha}.
\end{equation}
The isothermal conditions are:
\begin{equation}
\label{iso1}
\partial_lx^{\alpha}\partial_kx^{\alpha}=\frac{1}{2}\delta_{lk}(\partial_px^{\alpha})^2
\end{equation}
and one finds that $\delta A_{Dc}=\delta A_D$, given isothermality.  Note that the form of  $\delta A_{Dc}$, when isothermality is imposed, yields reparameterization invariance, since by definition reparameterization means that 
\begin{equation}
\label{reparam}
\epsilon_{ij}d\sigma_i\delta\sigma_j=0.
\end{equation}

Of course, isothermality is imposed only along a contour.  But since the isothermal relations are analytic in character, they extend from the contour to its interior.  In order that the contour integral for $A_{Dc}$ be useful, we must impose isothermality on the functions appearing in it.  A useful starting point in $d=3$ is the Enneper-Weierstrass construction, which reduces isothermality to an identity.

In $d=3$ the isothermal gauge conditions are imposed, as Enneper and Weierstrass showed long ago, by writing
\begin{equation}
\label{ew}
\partial_{\sigma}X=(1-f^2)g;\;\;\partial_{\sigma}Y=-i(1+f^2)g;\;\;\partial_{\sigma}Z=2fg
\end{equation}
where $g$  and $f^2g$ are meromorphic in the domain of $\sigma$ containing the minimal-surface contour (so $f$ can have poles if they are cancelled by zeroes of $g$).

The function $g$ changes under conformal transformations, and so for a wide class of contours, those conformally related to a standard contour, one can choose $g=1$.  Alternatively, for this class one can choose a standard contour in the $\sigma$ plane, for example, the unit circle, and determine an appropriate $g$.

Note that  two-dimensional areas are a trivial sub-case of the Weierstrass construction.  For simplicity, choose the $x^1$-$x^2$ plane as the $d=2$ space; then in the Weierstrass construction, take $f=0,g=1$. The anti-derivatives giving $x_1,x_2$ yield (up to a translation of the origin) $x_i=\sigma_i$, and the area formula of Eq. (\ref{extarea}) is in the non-contour form $\int d^2x$. This is, as already discussed, readily translated into a contour form.
Although in $d=2$ one chooses $f=0$  it is not necessary to take $g=1$.   But if $g\neq 1$ the specified contour in the $x_1,x_2$ plane is not the same as the contour in the $\sigma_1,\sigma_2$ plane.

   There are, of course, infinitely many contours which can be associated with an extremal area described by given  functions $f,g$, provided that the contours lie within the domain of harmonicity of these functions.  For any of these choices, the extremal area is given by:
\begin{equation}
\label{extarea}
A=\int d^2\sigma |g|^2(1+|f|^2)^2.
\end{equation}
 Weierstrass long ago showed\cite{weier} how, in principle, to solve the extremal-area problem for a simple non-intersecting closed contour composed of straight-line segments.  However, the solution has a character rather like the closely-related problem of Schwarz-Christoffel transformations, in which it is not at all easy to relate the parameters of the solution to the parameters of the contour.

\begin{figure}
\includegraphics[width=4in]{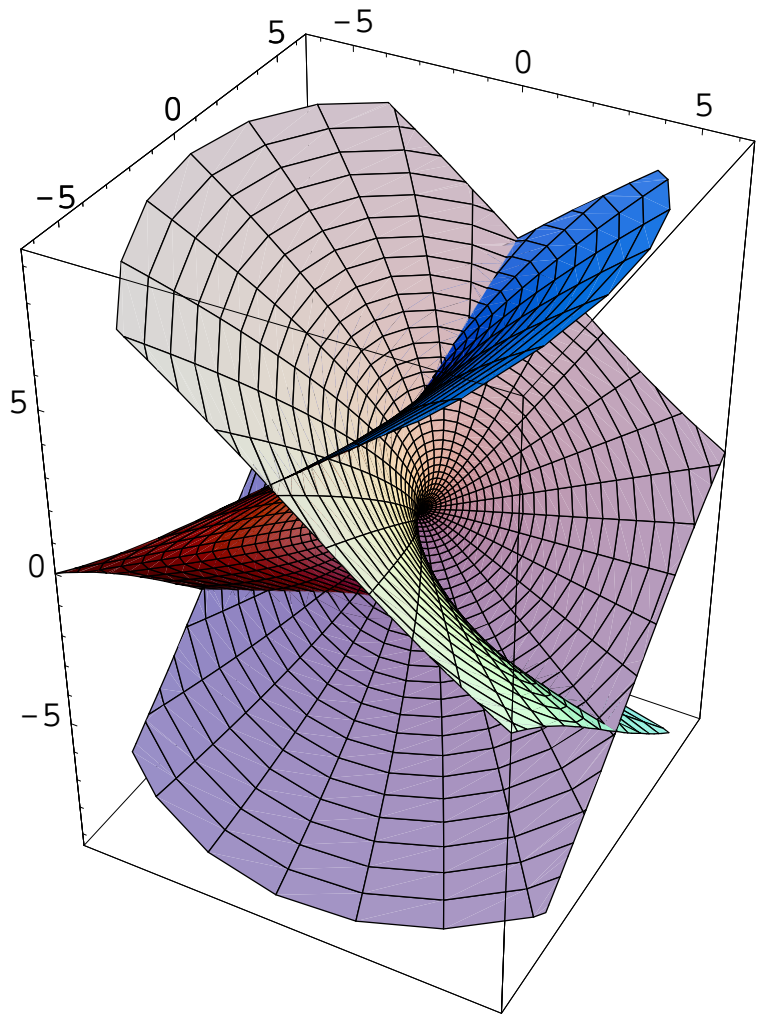}
\caption{\label{fig2} A portion of the Enneper surface.}
\end{figure}

Deviations from two-dimensionality might be profitably described with $f,g$ written as Taylor series.  Even a simple case such as keeping $g$=1 and saving only a linear term in $f$ (quadratic and cubic terms in the coordinates) results in a rather complicated surface known as Enneper's surface (see Fig. \ref{fig2}).

An alternative to using the Enneper-Weierstrass functions $f,g$, is expansion of the coordinates $X^{\alpha}=(X,Y,Z)$ themselves as Taylor series and then imposing isothermality.      We choose the origin of coordinates so that the surface passes through the origin at $\sigma =0$ and write:
\begin{equation}
\label{taylor}
X=\sum_1\frac{\alpha_N}{N}\sigma^N;\;Y=\sum_1\frac{\beta_N}{N}\sigma^N;\;Z=\sum_1\frac{\gamma_N}{N}\sigma^N.
\end{equation}  
One easily calculates the isothermality condition of Eq. (\ref{iso}) in terms of these coefficients:
\begin{equation}
\label{ident1}
0=\sum_1^N(\alpha_J\alpha_K+\beta_J\beta_K+\gamma_J\gamma_K)|_{J+K=N+1}
\end{equation}
It is straightforward to see that the $N=1$ term in $Z$ merely corresponds to a tilt of the surface, and can be set to zero.  Correspondingly, we set $\alpha_2,\;\beta_2=0$.  We can always choose $\alpha_1=i\beta_i=1$, as we did in $d=2$.  Under the assumption that all the $\gamma_N$ are of order of a smallness parameter $\epsilon$, there is a perturbative construction of isothermality:
\begin{equation}
\label{isopert}
\alpha_3-i\beta_3=-\frac{1}{2}\gamma_2^2;\;\alpha_4-i\beta_4=-\gamma_2\gamma_3;\;\alpha_5-i\beta_5=-\gamma_2\gamma_4-\frac{1}{2}(\alpha_3^2+\beta_3^2+\gamma_3^2),\dots
\end{equation}
One more condition can be imposed; the simplest is  to set the Enneper-Weierstrass function $g$ to unity, so that $\alpha_N+i\beta_N=0$ for $N\geq 2$.  The holomorphic function $f$ is just:
\begin{equation}
\label{fvalue}
f=\frac{1}{2}\partial_{\sigma}Z.
\end{equation}
Now we have just enough degrees of freedom for three contours, namely, two $\sigma$-space contours $\sigma_i(t)$ and the contour defined by ${\rm Re}Z(\sigma_i(t) )$, whhich can be chosen freely.  It is far from trivial, in general, to find explicitly the target-space contours $x^{\alpha}(t)$, although this can be done perturbatively.  One may say that this difficulty is the real difficulty of the Plateau problem, and we have nothing new to say about it.  But from the point of view of QCD dynamics, what is required is not to associate a set of target-space contours to areas, but to integrate over these contours, or areas, in a way consistent with external boundary conditions.  Integrating over the $\sigma$ contours and the $z$ contour will serve just as well, provided that no difficulties are encountered in imposing the boundary conditions.   

Let us define new functions $F_1,F_2$ as the real and imaginary parts of an antiderivative of $f^2$:
\begin{equation}
\label{deff}
F\equiv F_1+iF_2= -\int f^2=-\frac{1}{4}\int (\partial_{\sigma}Z)^2
\end{equation}
and then calculate the area from Eq. (\ref{finalarea}).  We specify a unique antiderivative by choosing $F(\sigma =0)=0$.   We assume that $f$ begins with a linear term and so $F$ begins with a quadratic term in $\sigma$ (a linear term in $F$ simply amounts to a tilt).

The area $A_{Dc}$, which we need no longer distinguish from the true extremal area $A$, is:
\begin{equation}
\label{totalarea}
A=\oint d\sigma_i[\frac{1}{2}\epsilon_{ij}\sigma_j+\frac{i}{2}(Z^*\partial_iZ-Z\partial_iZ^*)+\frac{i}{4}(F^*\partial_iF-F\partial_iF^*)].
\end{equation}

As advertised, there are $d=2$ constructions of a type already encountered  in Eq. (\ref{diri2}) or equivalently Eq. (\ref{2darea}), plus a quartic term. Note that the action (area plus kinetic terms) connected with these degrees of freedom is not a $d=2$ theory, but a $d=1$ field theory of the contours, although  these contours are coupled by a massless $d=2$ propagator, as in Eq. (\ref{2darea}).

The contour form of the Dirichlet area can be interpreted in terms of Abelian gauge potentials, as we now take up.

\subsection{Area and Abelian gauge potentials}

In any dimension $d>2$ the area can be expressed as a sum of $d$ contour integrals over $d$ Abelian gauge potentials, which are derived from the complex coordinates $X^{\alpha}$. Moreover, we can implement gauge transformations on these potentials and show, as expected, that they do not affect the area.  

In view of the area relations of Eq. (\ref{diri}, \ref{dirimulti}), we can express the area as an arbitrary linear combination of an $x$-form and a $q$-form and integrate by parts:
\begin{equation}
\label{arealambda}
A=\frac{1}{2}\oint d\sigma_i\epsilon_{ij}[\lambda x^{\alpha}\partial_jx^{\alpha}+(1-\lambda )q^{\alpha}\partial_jq^{\alpha}]
\end{equation}
where $\lambda$ is a numerical parameter.  (A different $\lambda$ can be used for each dimension, but we need not indicate that explicitly.) By using the Cauchy-Riemann relations one changes this to
\begin{equation}
\label{area5}
A=\sum_{\alpha}\oint d\sigma_iA^{\alpha}_i
\end{equation}
where
\begin{equation}
\label{gaugepot}
A^{\alpha}_i=\frac{1}{4}[q^{\alpha}\partial_ix^{\alpha}-x^{\alpha}\partial_iq^{\alpha}]+(\frac{1}{4}-\frac{\lambda}{2})\partial_i(x^{\alpha}q^{\alpha})\;\;{\rm (no\;sum)}
\end{equation}
One sees that a choice of $\lambda$ is equivalent to choosing a gauge for the $d$ gauge potentials $A^{\alpha}_i$;  we will work now in the gauge $\lambda =1/2$.  In this gauge we find that:
\begin{equation}
\label{currentform}
A^{\alpha}_i=\frac{i}{8}[(X^{\alpha})^*\partial_iX^{\alpha}- 
X^{\alpha}(\partial_iX^{\alpha})^*]\;\;{\rm (no\;sum)}.
\end{equation}
Another way of writing these $\lambda = 1/2$-gauge potentials is:
\begin{equation}
\label{complexform}
A^{\alpha}_i=\frac{1}{8}\epsilon_{ij}\partial_i|X^{\alpha}|^2\;\;{\rm (no\;sum)}.
\end{equation}

Straightforward calculation shows that 
\begin{equation}
\label{sumpot}
A^1_i+A^2_i=\frac{1}{2}\epsilon_{ij}\sigma_j+\frac{i}{2}[F^*\partial_iF-F\partial_iF^*]
\end{equation}
and, according to Eq. (\ref{currentform}), 
\begin{equation}
\label{zpot}
A^3_i=\frac{i}{2}[Z^*\partial_iZ-Z\partial_iZ^*].
\end{equation}
 The area constructed from the sum of the three gauge potentials is just that of Eq. (\ref{totalarea}).

 It is interesting to note that we can reinterpret the degrees of freedom in the area law of Eq. (\ref{totalarea}) as consisting of one Abelian gauge potential coupled to a complex scalar field $Z$, with quadratic and quartic terms in $\partial_{\sigma}Z$ ($Z$ can be expressed in terms of $\partial_{\sigma}Z$ by Eq. (\ref{antider})).  These are just the degrees of freedom in Witten's \cite{wit} description of spherically-symmetric $SU)2)$ gauge potentials, which he used to describe certain multi-instanton configurations.  Comtet \cite{comtet} has shown that the equations for extremal surfaces in $d=2+1$ can be reduced to Witten's instanton equations.  We will discuss this connection in detail elsewhere.

\section{Two-dimensional dynamics}
\label{appl}

 We begin with a description of the area law in Euclidean ($d=2$) space, and then, to set the stage for higher dimensions, complete the problem by considering $d=1+1$.

\subsection{${\bf d=2}$}

 For two spinless quenched quarks, described by a field $\phi (x)$, the Euclidean two-point mesonic Green's function is written:
\begin{equation}
\label{2point}
G(u-v)=\langle 0| \phi^{\dagger}\phi(v)\phi^{\dagger}\phi (u)|0\rangle =\int_0^{\infty}ds\int_0^{\infty}ds'
\int (dx)(dy)e^{-I}
\end{equation}
where $(dx)(dy)$ is a path integral over quark contours $x_i(\tau )$ and anti-quark contours $y_j(\tau ')$ and the action $I$ is:
\begin{equation}
\label{2daction}
I=\frac{M}{2}\int_0^{s}d\tau \dot{x}_i^2+\frac{M}{2}\int_0^{s'}d\tau '\dot{y}_j^2+\frac{M}{2}(s+s')+\frac{K_R}{2}\int_0^s d\tau \epsilon_{ij}x_i\dot{x}_j+\frac{K_R}{2}\int_0^{s'} d\tau ' \epsilon_{ij}y_i\dot{y}_j.
\end{equation}
In writing this action we have omitted self-mass terms, coming from quadratic contour integrals involving two quark (or two anti-quark) orbits.  These are necessary for insuring gauge invariance, and so must implicitly be understood to be included when questions of gauge invariance are brought up.
In Eq. (\ref{2daction}) $K_R$ is the string tension, $M$ is the quark mass, and the  boundary conditions on the path integral are:
\begin{equation}
\label{bc}
x_i(\tau =0)=u_i;\;x_i(\tau =s)=v_i;\;y_j(\tau '=0)=v_j;\;y_j(\tau '=s')=u_j.
\end{equation}
The mesonic spectrum is revealed by the poles in momentum of the Fourier transform of this Green's function.
 The two propagators in the Green's function are represented by the proper-time integrals.  In two-dimensional QCD $K_R=g^2C_R$, where $g^2$ is the coupling and $C_R$ is the quadratic Casimir eigenvalue.

Evidently the action is quadratic in the field variables, and thus exactly solvable.  However, one must be careful about the sign of the area term, since contours of both orientations occur.
This Euclidean action is simply the Minkowski action for a particle in a constant magnetic field, and the Green's function and particle spectrum are found from the classical solutions to the equations of motion.   The $M=0$ case can only be studied by postponing the limit until all calculations have been performed.

\begin{figure}
\includegraphics[width=4in]{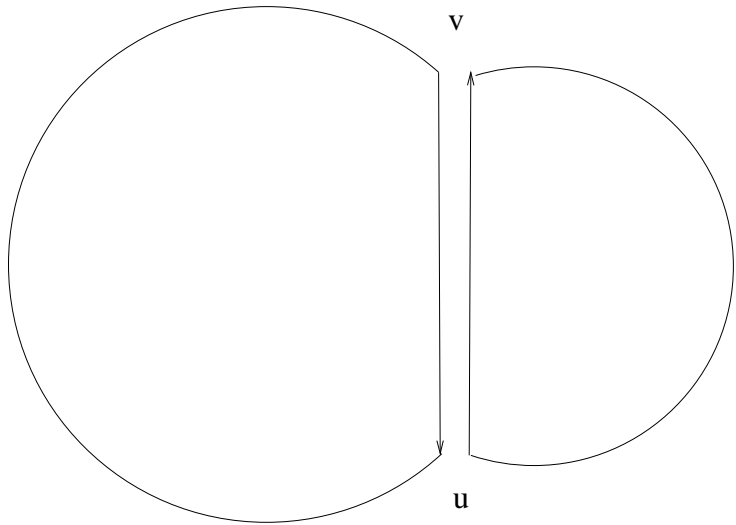}
\caption{\label{fig3} Geometric picture of the areas associated with the quark and the antiquark; each is bounded by a portion of a circle and a straight line between the initial and final positions of the $q\bar{q}$ pair.}  The contributions of the straight lines cancel in the action, so that the action depends on the contour of the overall area.
\end{figure}

The classical solution is conveniently written in terms of the complex variables
\begin{equation}
\label{complex2}
X=x_1+ix_2;\;U=u_1+iu_2;\;V=v_1+iv_2
\end{equation}
and is:
\begin{equation}
\label{classical}
X-U=(V-U)\frac{e^{-iK_R\tau /M}-1}{e^{-iK_Rs/M}-1}.
\end{equation}
The action for, say, the quark (proper time $s$) has the kinetic part
\begin{equation}
\label{kinetic}
\frac{M}{2}\int_0^sd\tau \dot{x}_i^2=(\frac{K_R^2s}{4M})\frac{(v_i-u_i)^2}{1-C}+\frac{Ms}{2}
\end{equation}
where in this and future equations we will use the shorthand
\begin{equation}
\label{sc}
C\equiv \cos (\frac{K_Rs}{M});\;S\equiv \sin (\frac{K_Rs}{M}). 
\end{equation}
The anti-quark contributes an identical kinetic term, with $s\rightarrow s'$.
The area part of the action for the quark is:
\begin{equation}
\label{areaaction}
\frac{K_R}{2}\int_0^s d\tau \epsilon_{ij}x_i\dot{x}_j=K_R[\frac{1}{2}(v_1u_2-v_2u_1)+\frac{(v_i-u_i)^2}{4(1-C)}(\frac{K_Rs}{M}-S)].
\end{equation}
This represents the sum of two areas.  The $v_1u_2-u_1v_2$ part is the area of a triangle whose vertices are the origin, $u_i$, and $v_i$, and the remainder is the area bounded by the quark contour going from $u_i$ to $v_i$ and a straight line from $v_i$ to $u_i$.
The anti-quark area term has the same form with two changes: 1) the sign of the $v_1u_2-v_2u_1$ (triangle) term is reversed; 2) $s$ is replaced by $s'$ in the remainder.  The sum of the quark and anti-quark terms is just the area bounded by the sum of the quark and anti-quark contours (see Fig. \ref{fig3}).  The total action, then, is of the form:
\begin{equation}
\label{itot}I=(\frac{K_R^2s}{4M})\frac{(v_i-u_i)^2}{1-C}+\frac{Ms}{2}+\frac{K_R(v_i-u_i)^2}{4(1-C)}(\frac{K_Rs}{M}-S)]+(s\leftrightarrow s').
\end{equation}
As required, this is translationally-invariant.

The area terms (last terms on the right in Eq. (\ref{itot})) have an interesting geometric interpretation. Elementary geometry shows that the quark area is that of a circle of radius
\begin{equation}
\label{circle}
R=\frac{|u-v|}{2|\sin (K_Rs/2M)|},
\end{equation}
less an area bounded by a portion of the boundary of the circle (the term linear in $s$)  and a chord drawn from $u_i$ to $v_i$ (the term $-S$).  This is shown in Fig. \ref{fig3}, for the case $\pi<K_Rs/M<2\pi$.  For $K_Rs/M$ in the range $(2\pi N, 2\pi (N+1)$ the circle is covered $N$ times.  It is, of course, this infinitely multiple covering of areas which allows the buildup of poles in the Fourier transform of the Green's function.

In forming the mesonic Green's function by integrating over proper times, one must be careful, because
the sign of the area term changes according to whether the quark is to the left or to the right of the anti-quark (as viewed with respect to the straight line joining $u$ and $v$).  This necessitates the introduction of absolute-value signs at various places, and complicates the analysis.  The same issues occur in $d=1+1$ dimensions, which we take up now. 

\subsection{${\bf d=1+1}$} 

First we must give the translation from Euclidean to Minkowskian variables.  In so doing, we must distinguish covariant from contravariant indices in the usual way, and we will use Greek indices for Minkowski-space variables.  The change from Euclidean to Minkowskian space in higher dimensions is entirely analogous to the $d=1+1$ discussion that we give next.

Since in the Euclidean version it was possible to identify the two target-space coordinates with the two surface parameters, this must also be possible in Minkowski space; consequently, the surface coordinates are Minkowskian, described as $\sigma^{0},\sigma^{1}$ with metric $(\sigma^{0})^2-(\sigma^{1})^2\equiv \sigma^2$.  Evidently the Euclidean area formula is to be replaced by the Minkowski-space formula
\begin{equation}
\label{minkarea}
A_m=\int d^2\sigma (-\eta )^{1/2}
\end{equation}
where $\eta$ is the determinant of the Minkowski-space world sheet metric, of signature (1,-1).  This ``area" is real, as it must be.  Of course, just as in $d=2$ we may identify the world sheet with the target space, and the area can be written in contour form as:
\begin{equation}
\label{contourmink}
A_m=\oint d\sigma^{\alpha}\epsilon_{\alpha \beta}x_{\lambda}\partial^{\beta}x^{\lambda}.
\end{equation}
Manipulations similar to those for $d=2$ show that the complete Minkowskian action $I_m$ for the $q\bar{q}$ system is, analogous to the Euclidean action of Eq. (\ref{2daction}),
\begin{equation}
\label{imink}
-I_m =\frac{M}{2}\int_0^{s}d\tau \dot{x}_{\alpha}\dot{x}^{\alpha}+\frac{M}{2}\int_0^{s'}d\tau '\dot{y}_{\beta}\dot{y}^{\beta}+\frac{M}{2}(s+s')+K_R\oint dx^{\alpha}\oint dy_{\alpha} \Delta (x-y). 
\end{equation}
The $q\bar{q}$ Green's function is constructed from the integral:
\begin{equation}
\label{greenmink}
\langle T(\phi (x)\bar{\phi}(y))\rangle =\int_0^{\infty}ds\int_0^{\infty}ds'e^{iI_m}.
\end{equation}
In Eq. (\ref{imink} the quantity $\Delta$ is an inverse of the d'Alembertian, and must be real in order to have a real area law in the action.  We cannot use the Feynman propagator $\Delta_F$, which has both real and imaginary parts:
\begin{equation}
\label{feynprop}
\Delta_F(x-y)=\frac{i}{4\pi}\ln |x|^2-\frac{1}{4}\theta (x^2).
\end{equation}
Instead we use only the real part, which is also the old Feynman-Wheeler propagator, half the sum of the retarded and advanced propagators.    So we keep (as in \cite{co77} in $d=3+1$) only the $\theta$-function, and identify
\begin{equation}
\label{realprop}
\Delta (x-y)=\frac{-1}{4}\theta [(x-y)^2].
\end{equation}
Actually, as \cite{co77} points out, this is gauge-equivalent to replacing the $\theta$-function by $-\theta [-(x-y)^2]$.

We need not give further details of this well-known $d=1+1$ problem which  is, of course, the standard hyperbolic problem of acceleration in a constant electric field of strength $K_R$, so that  sine and cosine in the $d=2$ formulas becomes sinh and cosh. The energy $E$ of the quark (or anti-quark) is constant, and given by:
\begin{equation}
\label{mink}
E=M\gamma +K_R|x^1-y^1|
\end{equation} 
where the coordinate $x^1(y^1)$ is the quark (anti-quark) spatial variable, and $\gamma$ is the usual relativistic factor for either the quark or the anti-quark (which always have equal and opposite velocities).  The absolute value sign on $x^1-y^1$ is the Minkowski analog of the signs we mentioned for the Euclidean case, and occurs because the field strength changes sign as the quark and anti-quark pass each other. WKB quantization leads to the standard formula $E^2=\pi K_R N$ for large integral $N$ (in which limit the mass $M$ can be neglected).

\section{Three-dimensional dynamics}
\label{3dsec}

 First we study the $d=3+1$ problem of $q\bar{q}$ bound states at large angular momentum $\ell$, which at the classical level is actually a $d=2+1$ problem because the orbits are planar, and compare it to two earlier approaches to the same problem.  Then we give a scaling law for corrections to the classical result arising from certain quantum fluctuations.

We are interested, as in the previous section, in the bound-state masses of $q\bar{q}$ systems considered classically.  Of particular interest is the limit of zero quark mass $M$, where we will find a novel effect in which the quark kinetic terms in the action vanish in the limit.  In actuality, even massless quarks protected by chiral symmetry acquire a constituent mass, since confinement breaks chiral symmetry, but we will ignore these effects here to make our main point, which concerns the asymptotic slope of the QCD linearly-rising Regge trajectory.  We will, as in the $d=1+1$ case, ignore explicit mass-generation terms involving two quark or two anti-quark contours, but these are important for maintaining gauge invariance.  

Classical orbits are sensible for large angular momentum $\ell$, which is all that we will study here.  In this limit, the $q\bar{q}$ orbits lie in a spatial plane (and, of course, evolve in time), with the result that one spatial coordinate can be ignored and so a classical problem in $d=3+1$ can be formulated in $d=2+1$    Minkowski space.  We expect massless quarks to get constituent masses of order $K_F^{1/2}$, and this effect leads to corrections to our results which vanish at large $\ell$.

There are several works from long ago that have also considered this problem, and we will be concerned with comparing the results found here to two of them \cite{cktwz,co77}.  In both these two works and in the present investigation the bound-state masses lie on a linearly-rising Regge trajectory in the massless quark limit.  It turns out that in this limit the Regge trajectory has slope $\alpha'(0)$ obeying the usual Veneziano/string relation $\alpha'(0)K_F=1/(2\pi )$ for both the extremal-area picture and for Ref. \cite{cktwz}, which is based on a naive string picture.  However, the third model \cite{co77}, which is based on the effective propagator of Eq. (\ref{gluprop}), has a different and smaller value for this product.  

At the classical level, at least, the present work is virtually identical in its predictions to that of Ref. \cite{cktwz}.  In fact, the extremal-area action is identical to an action (not explicitly quoted in \cite{cktwz}) which yields the results of that reference.  But the present work and any form of string theory must differ at the quantum level, as we have discussed earlier.

\subsection{The effective-propagator model}

Consider first the effective-propagator model of Ref. \cite{co77}.  In this model the action corresponding to Eq. (\ref{imink}) is:
\begin{equation}
\label{propaction}
-I_m=\frac{M}{2}\int_0^{s}d\tau \dot{x}_{\alpha}\dot{x}^{\alpha}+\frac{M}{2}\int_0^{s'}d\tau '\dot{y}_{\beta}\dot{y}^{\beta}+\frac{M}{2}(s+s')+\frac{K_F}{2}\int_0^sd\tau \int_0^{s'}d\tau' \dot{x}^{\alpha}(\tau )\dot{y}_{\alpha}(\tau')\theta [-(x-y)^2]
\end{equation}
where only the real part of the propagator has been saved.  (We study here only the case of equal mass $M$ for the quark and anti-quark; the general case can be found in \cite{co77}.)  

The contours chosen in \cite{co77} to solve the classical equations of motion in $d=3+1$ for the effective-propagator dynamics, with action as in Eq. (\ref{propaction}), are:
\begin{eqnarray}
\label{contour77}
x^{\alpha} & = & (\frac{t\tau}{s},\;\rho \cos (\frac{\phi\tau}{s}),\;\rho \sin (\frac{\phi\tau}{s}),\;0);\\ \nonumber
y^{\alpha} & = & (\frac{t'\tau'}{s'},\;-\rho \cos (\frac{\phi'\tau'}{s'}),-\;\rho \sin (\frac{\phi'\tau}{s'}),\;0).
\end{eqnarray}
Here $\rho$ is the common radius of the quark or anti-quark orbit, and parameters are chosen so that at $\tau = s$ one has $x^{\alpha}=(t,\;\rho \cos \phi,\;\rho \sin \phi,\;0)$ (and similarly for the anti-quark).  For the purposes of finding bound-state energies, boundary conditions at $\tau=0$ are irrelevant.  The $z$-coordinate is fixed, and so the problem is really in $d=2+1$.  From now on we omit writing the inert $z$ coordinate.

As explained in \cite{co77}, to calculate the total energy $E$ of and angular momentum $\ell$ of a bound state one can set $s=s',\;t=t',\;\phi =\phi'$. These prescriptions are consistent with the requirement that we are working in the bound-state center of mass, so that the energy $E$ is actually the mass of the bound state. We then impose four conditions  corresponding to the four external variables of the problem:  $s,t,\rho ,\phi$.    Following the imposition of these conditions we introduce a derived quantity $\omega$, obeying $\phi=\omega t$ where $\omega$ is the constant angular velocity of the quark or anti-quark.  The common quark linear velocity is   $v=\rho \omega$.  

First, a steepest-descent integration over these proper time variables in the Green's function shows that $s=t/\gamma,\;s'=t'/\gamma$ where $\gamma =(1-v^2)^{-1/2}$  is the usual relativistic factor. (We will see that only the kinetic terms depend on the proper times.)    The second and third conditions follow from the definitions of energy and angular momentum as derivatives of the action:
\begin{equation}
\label{energy}
E=-\frac{\partial I_m}{\partial t};\;\ell = \frac{\partial I_m}{ \partial \phi}.
\end{equation}
  The fourth condition comes from the   equations of motion.  There is an azimuthal component of these equations which is identically satisfied because the propagator we use is half the sum of the advanced and retarded propagators.  The radial component of the equations of motion is perhaps most easily found from the equation
\begin{equation}
\label{radmotion}
0=\frac{\partial E}{\partial \rho}|_{\ell},
\end{equation}
which is a way of saying that $\dot{\rho}=0$.
 From these four conditions one eliminates $s,M\gamma,$ and $\omega$ to find the energy as a function of $\ell$; this is, not unexpectedly, a linearly-rising Regge trajectory. 

For  the effective-propagator model the action is (ignoring terms which vanish as $t\rightarrow \infty$):
\begin{equation}
\label{glupropaction}
-I_m=\frac{M}{2}[\frac{t^2-\rho^2\phi^2}{s}+s]+\frac{M}{2}[\frac{t^2-\rho^2\phi^2}{s'}+s']+K_F[t^2\xi +\rho^2\phi \sin (\xi \phi )].
\end{equation}
Here $\xi$ is the positive root of the equation
\begin{equation}
\label{etaeqn}
t^2\xi^2=2\rho^2(1+\cos (\xi \phi)).
\end{equation}
The meaning of $\xi$ is that it is the value of $|\tau -\tau'|/s$ at which the $\theta$-function in the effective-propagator action of Eq. (\ref{newlaw}) vanishes.
A little algebra shows that one may write $\xi =\theta /\omega t$ where $\theta$ is the positive root of
\begin{equation}
\label{thetaeq}
\theta^2 = 2v^2(1+\cos \theta ).
\end{equation}
For massless quark and anti-quark, with $v=1$, one finds \cite{co77} $\theta=1.478$, and in the non-relativistic limit of small $v$, $\theta=2v$.  

It remains to carry out the action derivatives in Eqs. (\ref{energy}).  In so doing, the actions is expressed as a function of $s,\rho,\phi,t$.  After the derivatives are taken, one may then substitute $s=t/\gamma,\phi =\omega t$.  We are only interested in the relativistic (massless) case, so where it is safe to do so we set $v=1$ in the resulting equations.  This may not be done in the product $M\gamma$ until this combination has been eliminated among the three equations.  We find:
\begin{equation}
\label{3equations}
E=2M\gamma +\frac{K_F}{\omega}C_1;\;\ell =\frac{2M\gamma}{\omega}+\frac{K_F}{\omega^2}C_2;\;M\gamma \omega =K_FC_3
\end{equation} 
where the $C_i$ are the values of functions $C_i(v)$ at $v=1$; these values depend on $\theta (v=1)$.  A couple of lines of algebra yields a linearly-rising Regge trajectory:
\begin{equation}
\label{regge}
E^2=C_4K_F\ell;\;C_4=\frac{(2C_3+C_1)^2}{2C_3+C_2}.
\end{equation} 

\subsection{The extremal-area model}

For the extremal-area action, the only difference is that the area-law interaction term $I_{A}$ has a $\theta$-function with a different argument as prescribed in Eq. (\ref{finalarea}) (when transcribed to Minkowski space):
\begin{equation}
\label{newlaw}
-I_A=\frac{K_F}{2}\int_0^sd\tau \int_0^{s'}d\tau' \dot{x}^{\alpha}(\tau )\dot{y}_{\alpha}(\tau')\theta [-(\sigma -\sigma')^2]
\end{equation}
where $\sigma (\tau )\;(\sigma (\tau'))$ is the quark (anti-quark) contour on the world sheet, and $x^{\alpha}(\tau )\equiv x^{\alpha}[\sigma (\tau )]$ is the target-space quark contour.    In the non-relativistic limit (where $\theta = 2v$ for the effective-propagator model) the extremal-area model and the effective-propagator model agree.    In the massless limit where $\rho \omega =v=1$, and using $\phi =\omega t$, the area part of the action is:
\begin{equation}
\label{massless}
-I_{mA}|_{v=1}=\frac{\pi \rho t}{2}
\end{equation}
which is of the expected form.  

Note that the orbits for the effective-propagator model  (Eq. (\ref{contour77})) are helices, and taken together they form a double helix.  These  helices form the boundaries of a well-known extremal area, the helicoid, and it is not surprising that the extremal-area model has a solution involving this particular extremal surface.  The helicoid has an axis running along the proper-time direction. The contour $z^{\alpha}(\sigma^0 ,\sigma^1 )$ spanned by the minimal surface can be very simply expressed in isothermal coordinates:
\begin{equation}
\label{helicoid}
z^{\alpha}=(\frac{t\sigma^0}{s},\;\frac{t}{\phi}\sin (\frac{\phi\sigma^1}{s})\cos (\frac{\phi \sigma^0}{s}),\;\frac{t}{\phi}\sin (\frac{\phi\sigma^1}{s})\sin (\frac{\phi \sigma^0}{s})).
\end{equation}
 It is elementary to check that $z^{\alpha}(\sigma^0,\sigma^1)$ is harmonic and isothermal.

The next step is to relate the Wilson-loop contour as expressed in world-sheet variables to the kinematic variables.  For the quark orbit we identify the proper time $\tau$ with $\sigma^0$, and for the anti-quark we identify $\tau'$ with $\sigma^0$.  The other world-sheet variable $\sigma^1$ is independent of proper time, and takes on values differing by a sign for the quark and anti-quark.  We have: 
\begin{equation}
\label{helices}
\sigma^0=\tau,\;\frac{\rho \phi}{t} =\sin (\frac{\phi\sigma^1}{s})\;{\rm (quark\;orbit)};\;\sigma^0=\tau',\;\frac{\rho\phi}{t} =-\sin (\frac{\phi\sigma^1}{s})\;{\rm (anti-quark\;orbit)}.
\end{equation}
The resulting  helical orbits are essentially those of Eq. (\ref{contour77}) used for the effective-propagator action.

The real difference between the effective-propagator model and the extremal-area model now shows up.  Because $\rho \phi /t=v$, where $v$ is the linear velocity of the quark or anti-quark, we can write for the boundary values of $\sigma^1$ on the contour:
\begin{equation}
\label{sigma1}
M\rightarrow 0:\;\;\frac{\phi \sigma^1}{s}=\pm \arcsin v \rightarrow \pm \frac{\pi}{2}.
\end{equation}
One now sees, by consulting the argument of the $\theta$-function in the extremal-area action of Eq. (\ref{newlaw}) and comparing it to the $\theta$-function of the effective-propagator action of Eq. (\ref{propaction}), that the effective limiting value of the angle $\theta$ introduced in Eq. (\ref{thetaeq}) is $\pi$ at zero quark mass, or unit quark velocity.  This angle $\theta$, as discussed in \cite{co77}, measures the retardation (or advance) of signals exchanged between quark and anti-quark.  The rate of approach to the limit is such that
\begin{equation}
\label{newtheta}
\sin \theta \rightarrow \frac{ {\rm const.}}{\gamma}
\end{equation}
(we do not need to know the constant).  Some straightforward algebra shows that at $\theta=\pi$ the equation of motion is different in structure from that of the effective-propagator model, as given in the third equation of Eq. (\ref{3equations}), since the coefficient of $K_F$ vanishes proportional to $\sin \theta$.  Given the rate of approach of vanishing, the equation of motion for the extremal-area model is:
\begin{equation}
\label{newmotion}
M\gamma \omega = \frac{C'_3K_F}{\gamma}
\end{equation}
for some (irrelevant) constant $C'_3$.  This is easy to solve:
\begin{equation}
\label{newgamma}
M\rightarrow 0:\;M\gamma \rightarrow (M\rho K_FC'_3)^{1/2}\rightarrow 0.
\end{equation}
The kinetic terms drop out of the action, leaving only the area term, in the massless limit.

Of course, if the massless limit were in fact attainable, and in this limit the quark kinetic energy vanished, classical calculations would be highly suspect.  An infrared massless particle does not possess the kind of localization expected for classical physics to be valid.  In fact, a constituent mass of $O(K_F^{1/2})$ will be generated, so there is some localization.  We should require that $M\gamma\rho \geq 1$ as a minimum requirement of localizability.  This means, using Eq. (\ref{newgamma}) and the easily-derived scaling $\rho \sim (\ell /K_F)^{1/2}$, which holds in the massless limit, that:
\begin{equation}
\label{mrho}
M\gamma \sim (M\rho K_F)^{1/2}>\frac{1}{\rho};\;\rho^3\sim (\frac{\ell}{K_F})^{3/2}>\frac{1}{MK_F}
\end{equation}
which, with $M\sim K_F^{1/2}$, simply means large $\ell$.  So we will continue to use the massless limit as a formal device, expecting corrections from real masses which vanish at large $\ell$.

 This extremal-area action is:
\begin{equation}
\label{minkarea1}
-I_{mA}\equiv K_F\int d^2\sigma (-\eta )^{1/2}= \frac{K_Ft^2}{\phi}\{\arcsin (\frac{\rho \phi}{t})+\frac{\rho \phi}{t}[1-(\frac{\rho \phi}{t})^2]^{1/2}\}.
\end{equation}
For comparison with the work of Callan {\it et al} \cite{cktwz} we note that this action has the integral representation:
\begin{equation}
\label{calaction}
-I_{mA}=2K_F\int_0^{\rho}dR[t^2-(R\phi )^2]^{1/2}.
\end{equation}

\subsection{Numerical values for the three models}

The constants for the effective-propagator case were worked out long ago \cite{co77} and yield
\begin{equation}
\label{effprop}   
E^2=9.91K_F\ell\;{\rm (effective\;propagator)}.
\end{equation}
This yields a Regge slope--string tension product $\alpha'(0)K_F$ of about 0.101, rather small compared to the Veneziano/string theory product of $1/(2\pi )$=0.159.

For the extremal-area model in leading order in $1/\gamma$, we find $C_1=\pi$, $C_2=\pi /2$, and to all intents and purposes $C_3=0$, because of the extra factor of $1/\gamma$ in Eq. (\ref{newmotion}).  The values of $C_1,C_2$ are the same as found in the elementary string model of Callan {\it et al} \cite{cktwz}.  (This reference does not quote a value for $C_3$.)  Mathematically, at least, this is no coincidence;  differentiation of the action for the extremal-area model as given in Eq. (\ref{calaction}) yields precisely the integrals for $E,\ell$ of Ref. \cite{cktwz}.   But the physical interpretation is presumably quite different, since a string theory differs at the quantum level from the extremal-area theory.
The values of the $C_i$ of the extremal-area model yield, from Eq. (\ref{regge}), the usual Veneziano/string relation: 
\begin{equation}
\label{evsl}
E^2=2\pi K_F\ell\;{\rm (extremal\;area)}.
\end{equation}

One might expect that corrections to these leading-order results are of $O(\gamma^{-1}) \sim \ell^{-1/4}$, in view of 
Eq. (\ref{newgamma}) and the leading-order scaling $\rho \sim (\ell /K_F)^{1/2}$.  In fact, because of the specific form of the area-law action in Eq. (\ref{minkarea1}), determined in part by dimensional considerations, it turns out that the leading corrections are $O(\gamma^{-2})\sim \ell^{-1/2}$, leading to:
\begin{equation}
\label{scaling}
E=(2\pi K_F \ell)^{1/2}[1+O(\frac{M}{(K_F\ell )^{1/2}})];
\end{equation}
that is, the leading correction is just a mass term.  It appears that at least some quantum corrections coming from orbital fluctuation in the (classically irrelevant) $z$-direction have the same scaling, as we discuss next.

\subsection{Scaling laws at large angular momentum from quantum fluctuations}

We estimate contributions from $z$-fluctuations away from planar circular orbits, as expresssed by the quartic terms in the area of Eq. (\ref{totalarea}).  These estimates will be done in Euclidean space, which should have no effect on simple scaling laws.  The final result identifies
 a smallness parameter $\epsilon$ related to these deviations  from planarity:  
\begin{equation}
\label{scale}
\epsilon \sim (K_F\rho^2)^{-1/2}\sim \ell^{-1/2}.
\end{equation}

We will   take into account only the $\gamma_2$ term in the Taylor series expansion for $Z$ (Eq. (\ref{taylor})) to estimate the corrections.    Keeping only the $\gamma_2$ term corresponds to $F=-\gamma_2^2\sigma^3/12$ in Eq. (\ref{deff}).  From Eq. (\ref{totalarea}) for the area, multiplied by $K_F$ to find the area part of the action, one finds that the quartic contribution $I_4$ to the action  scales as:
\begin{equation}
\label{i4}
I_4\sim K_F\rho^6|\gamma_2|^4.
\end{equation}
The quadratic contributions to the action are all positive, and so the functional integral over $\gamma_2$ leads to an estimate of the typical value of $|\gamma_2|$:
\begin{equation}
\label{gamma2}
|\gamma_2|\sim (K_F\rho^6)^{-1/4}.
\end{equation}
Then $Z\sim \rho\epsilon^{1/2}$ and the area part of the action (which is dominant over kinetic terms) coming from the $F$ terms in Eq. (\ref{totalarea}) scales as:
\begin{equation}
\label{scale2}
I_z\sim K_Fz^2\sim K_F\rho^2\epsilon .
\end{equation}
To relate this scaling to the angular momentum $\ell$, we have already seen that the mass $M_{\ell}$ of a relativistic bound state with large $\ell$ scales as:
\begin{equation}
\label{scale3}
M_{\ell}\sim (K_F\ell )^{1/2}\sim K_F\rho,
\end{equation}
and that:
\begin{equation}
\label{scale4}
\rho \sim (\frac{\ell}{K_F})^{1/2};\;\epsilon \sim \ell^{-1/2}.
\end{equation}
Finally, the correction to the bound-state mass behaves like:
\begin{equation}
\label{correct}
M_{\ell}\sim (K_F\ell )^{1/2}[1+O(\ell^{-1/2}]=(K_F\ell )^{1/2}+O(K_F^{1/2}).
\end{equation}
More accurate estimates will be given elsewhere.

\section{Summary}
\label{sum}

In this paper we argued in some detail for expressing the dynamical area law of QCD as a law of extremal surfaces, based on the topological mechanism of confinement in center-vortex models.  In principle the area of an extremal surface can be expressed as a functional of the contour (Wilson loop) spanned by the surface.  Confined-particle dynamics and bound states follow from functional integrals over these contours, which amounts to a $d=1$ field theory.  This is unlike earlier attempts to deal with area-law dynamics, in which string surfaces associated with tubes of confined flux have  $d=2$ fluctuations which must be integrated over, even for a fixed Wilson loop.   These fluctuations, existing over the whole world surface of the flux tube, lead to severe complications \cite{polstrom}.  But in the center-vortex picture, fluctuations have effect only on subdominant (perimeter) terms and are much more easily dealt with.

We also pointed out some of the challenges associated with this law, such as the existence of several extremal areas for certain contours, and discussed some questions of the behavior of extremal areas for the sort of non-differentiable contours likely to occur in the functional integral over Wilson-loop contours.  Certain more complicated contour configurations may have to do with adjoint string breaking and related phenomena.   

The contour form expression for an extremal area has an interpretation in terms of Wilson loops of $d$ Abelian gauge potentials, in $d$ dimensions.  There is a natural definition of a gauge transformation, leaving the area unchanged, on each of these gauge potentials.  It is interesting to note that in $d=3$ the degrees of freedom can be reinterpreted as one Abelian gauge potential plus a complex scalar field, which happen to be the degrees of freedom identified by Witten \cite{wit} in his spherical {\it ansatz} for $SU(2)$ gauge fields.  Comtet \cite{comtet} has shown that the $SU(2)$ self-duality equations for these degrees of freedom are precisely those  extremalize a surface in $d=2+1$.  What the  connection is between our work and this earlier work remains to be seen; it is under investigation.

These contour formulas are simple generalizations of area formulas in $d=2$, which express an area as the interaction of a massless $d=2$ gauge propagator with currents running along the Wilson-loop contour, just as required by $d=2$ QCD.  However, their simplicity is deceiving, since the contour form of the area law holds only for the Dirichlet integral, which expresses a true extremal area only under the imposition of isothermal gauge conditions.  Isothermality leads to expressions both quadratic and quartic in the contours, and so the extremal-surface form is an interacting field theory of these $d=1$ contours. 
 
We gave one simple example in $d=3+1$ (actually applicable to $d=3+1$ problems) of a classical extremal-area law, in which the quark and anti-quark contours were helices and the extremal area a helicoid.  
We compared the large-$\ell$ results of this center-vortex model with two earlier models, one an effective-propagator model \cite{co77} and one a naive string model \cite{cktwz}, and found that the present results differ from those of \cite{co77} but agree with those of \cite{cktwz}, even to the extent that the actions of \cite{cktwz} and of the present paper are mathematically identical.  However, this does not mean that the {\em physical} content of these two models are the same; presumably at the quantum level the string model would involve functional integrations over string fluctuations, while in the center-vortex picture such fluctuations are missing.

In the heavy-quark limit, all approaches including that given here agree with one another and with the conventional quantum mechanics of a linearly-rising potential.  But there is a curious effect in the limit of  quarks whose mass $M$ approaches zero.  One generally expects that  in this limit the quark kinetic energy $M\gamma$, where $\gamma$ is the usual relativistic factor, is finite.  This is, for example, what happens in the effective-propagator model.  But in the extremal-area model, retardation effects lead to an extra factor of $1/\gamma$ in the force term of the $q\bar{q}$ equation of motion, which has the consequence that $M\gamma\sim M^{1/2}$ in the limit, and thus vanishes.  The quark kinetic terms in the action no longer contribute, and only the area-law part is left.  This, of course, is what string theory would prescribe, and in fact the extremal-area model studied here yields precisely the Veneziano/string result $\alpha'(0)K_F=1/(2\pi )$.  Generally, in any model where quark kinetic terms survive the massless limit, this product is smaller than $1/(2\pi )$.    

Under the circumstances where the semi-classical developments of this paper are applicable, such as for bound states of large angular momentum $\ell$, there is a smallness parameter $\ell^{-1/2}$ which allows one to expand deviations from three-dimensional contours, dominant at large $\ell$, into the fourth spatial dimension.  We made some estimates which showed that such fluctuations yield corrections vanishing at large $\ell$, as expected.

The dynamics of extremal surfaces has just begun to be explored, and future work will probe more deeply into several of these matters.  

\begin{acknowledgments}
I am happy to thank Per Kraus and Shmuel Nussinov for enlightening conversations.
\end{acknowledgments}

\end{document}